# Fundamental limitations of half-metallicicity in spintronic materials

A. Solontsov[1,2,*]

[1]N.L. Dukhov Research Institute for Automatics, 22 str. Suschevskaya, Moscow 127055, Russia,

[2]State Center for Condensed Matter Physics, 6/3 str. M. Zakharova, Moscow 115569, Russia
Dated: October 27, 2013

Zero-point spin fluctuations are shown to strongly influence the ground state of ferromagnetic metals and to impose limitations for the fully spin polarized state assumed in half-metallic ferromagnets, which may influence their applications in spintronics. This phenomenon leads to the low-frequency Stoner excitations and cause strong damping and softening of magnons in magnetoresistive manganites observed experimentally.

**Introduction**

The complete spin polarization of itinerant electrons in the ground state of half-metallic ferromagnets is of essential interest for spin electronics [1]. This phenomenon is believed to be realized in magnetoresistive manganites, which are canonically described within the double-exchange (DE) model [2] leading in the mean-field approximation to the $P=100\%$ polarization. However, the tunnel junctions experiment lead to the essentially less values of the polarization [3,4] $P \sim 80\%$. This decrease in polarization is likely to be caused by mixing of electronic spins due to the effects of zero-point effects associated with magnons, phonons and spin fluctuations (SF) coupled to the magnetic system and giving rise to the minority spin electrons at the Fermi surface at zero temperature. The same effect at finite temperatures was discussed in Refs. 5 and 6, where it was associated with thermal excitations of magnons, phonons and SF. In other words, 100% polarization in the half-metallic ferromagnets can never be realized due to fundamental limitations related to zero-point effects and thermal excitations both in the ground state and at finite temperatures.

However, the lack of full polarization in half-metallic magnets at zero temperature up to now was out of the scope of investigations of spintronic materials. It can be also emphasized that due to the giant character of zero-point SF [7] their influence on the magnetic ground state and zero-temperature polarization seems to be the most important. However, the theory of SF accounting for both zero-point and thermal SF, and spin anharmonicity is lacking. In the present paper we formulate the generalized theory of zero-point and thermal SF and analyze the magnetic ground state and half-metallicity of ferromagnets.

**Model for spin fluctuations.**

To account for SF in the ground state of various types of ferromagnets containing itinerant electrons (not only itinerant electron magnets) we shall use a phenomenological Ginzburg-Landau (G-L) approach basing on the following effective Hamiltonian and time-dependent equations

$$\hat{H}_{eff} = \frac{1}{2}\sum_{\mathbf{k}} \chi_0^{-1}(\mathbf{k})|\mathbf{M}(\mathbf{k})|^2 + \frac{\gamma_0}{4} \sum_{\mathbf{k}_1+\mathbf{k}_2+\mathbf{k}_3+\mathbf{k}_4=0} (\mathbf{M}(\mathbf{k}_1)\mathbf{M}(\mathbf{k}_2))(\mathbf{M}(\mathbf{k}_3)\mathbf{M}(\mathbf{k}_4))$$

$$\frac{1}{\Gamma_0(\mathbf{k})}\frac{\partial \mathbf{M}(\mathbf{k})}{\partial t} = -\frac{\delta \hat{H}_{eff}}{\delta \mathbf{M}(-\mathbf{k})} \quad (1)$$

which describes interaction of SF and their dynamics. Here $\mathbf{M}(\mathbf{k},t) = \mathbf{M}\delta_{\mathbf{k},0} + \mathbf{m}(\mathbf{k},t)$ is the time-dependent order parameter, $\mathbf{M}$ is the magnetization, $\mathbf{m}(\mathbf{k},t)$ accounts for SF, $\chi_0(\mathbf{k})$ is the static paramagnetic susceptibility not accounting for SF, $\gamma_0$ is the bare mode-mode coupling constant not affected by SF, and $\Gamma_0(\mathbf{k})$ is the relaxation rate.

GL Hamiltonian (1) is usually treated as an expansion in terms of SF amplitudes which are considered to be small. Here we use the model based on the Hamiltonian (1) not assuming expansions in terms of SF amplitudes, which according both to the experiment [7] and theory [8] may be giant. We hope that this model accounts for the main features of strongly coupled zero-point and thermal SF.

The model (1) should be accompanied by the fluctuation-dissipation theorem relating the squared local magnetic moment (averaged SF amplitude) $M_L^2$ with dynamical magnetic susceptibilities $\chi_\nu(\mathbf{k},\omega)$,

$$M_L^2 = 2\hbar \sum_\nu \sum_{\mathbf{k},\omega} \operatorname{Im} \chi_\nu(\mathbf{k},\omega), \quad (2)$$

where $\nu$ denotes the transverse ($t$) and longitudinal ($l$) polarization, $\omega$ is the frequency of SF, $\sum_\omega = \int_0^\infty (d\omega/2\pi)$. Below we shall use the following phenomenological form for the dynamical susceptibilities

$$\chi_\nu^{-1}(\mathbf{k},\omega) = \chi_\nu^{-1} + c(\mathbf{k}) - i\frac{\omega}{\Gamma_0(\mathbf{k})} \quad (3)$$

supported both theoretically and experimentally [9], where $\chi_\nu$ are static susceptibilities, $c(\mathbf{k})$ accounts for their spatial dispersion, and $\Gamma_0(\mathbf{k})$ is the relaxation rate. Concentrating here on the ground state properties, we take into account only zero-point SF and neglect thermal ones.

One of the most important characteristics of the model (1) is the dimensionless spin anharmonicity parameter [8]

$$g_{SF} = \frac{2}{3}\gamma_0 \hbar \sum_\nu \sum_{\mathbf{k},\omega} \operatorname{Im} \chi_\nu^2(\mathbf{k},\omega), \qquad (4)$$

which is analogous to the arising in the theory of anharmonic crystals. Elementary estimates of $M_L$ and $g_{SF}$ show that the local magnetic moment due to zero-point SF is of the order of the number of itinerant electrons (in Bohr magnetons) per unit cell and the anharmonicity parameter is of order unity. So, SF in the ground state of magnetic metals should be considered as giant (which was supported experimentally [7]) and strongly interacting, which exclude their treatment within any perturbation scheme.

To calculate the free energy $F(M)$ we use the following set of integro-differential equations

$$F(M) = F_0(M) + \Delta F_{SF},$$
$$F_0(M) = \frac{1}{2\chi_0} M^2 + \frac{\gamma_0}{4} M^4, \qquad (5)$$
$$\Delta F_{SF}\{\chi_\nu(\mathbf{k},\omega)\} = \frac{1}{2}\int d\left(\chi_0^{-1}\right) M_L^2 = \hbar \sum_\nu \sum_{\mathbf{k},\omega} \int d\left(\chi_0^{-1}\right) \operatorname{Im} \chi_\nu(\mathbf{k},\omega),$$

where $F_0(M)$ is the Landau free energy not affected by SF, $\chi_0 = \chi_0(\mathbf{k}=0)$, $\Delta F_{SF}\{\chi_\nu(\mathbf{k},\omega)\}$ is the SF contribution. In the dynamical susceptibilities (3) the SF effects on the spatial dispersion $c(\mathbf{k})$ and relaxation $\Gamma_0(\mathbf{k})$ are vanishing in the ground state [10] but very important for the static susceptibilities, which we calculate self-consistently using thermodynamic relations

$$\chi_t^{-1} = \frac{1}{M}\frac{\partial F}{\partial M}, \quad \chi_l^{-1} = \frac{\partial^2 F}{\partial M^2} \qquad (6)$$

To find the analytical solution of equations (5) one must calculate the SF contribution to the free energy $\Delta F_{SF}$ as a function of static susceptibilities $\chi_\nu$. Following the ideas of the soft-mode theory of SF [8] we assume that the spatial dispersion of the inverse susceptibilities is strong enough, so that the parameter $\chi_\nu^{-1}/c(\mathbf{k}_B) \ll 1$ is small (where $\mathbf{k}_B$ is the wavevector at the Brillouin zone boundary) and expand the SF free energy $F(M)$ in powers of $\chi_\nu^{-1}$, This allows to solve (5) and leads to the free energy in the Landau form with the renormalized coefficients $\chi_0 \to \chi$, $\gamma_0 \to \gamma$, $g_0 \to g$,

$$\chi^{-1} = (1-5g)\chi_0^{-1} + \frac{5}{3}\gamma M_{L0}^2 < 0, \qquad (7)$$

$$\frac{\gamma}{\gamma_0} = \frac{g}{g_0} = \frac{1-5g}{1+6g}, \tag{8}$$

where $M_{L0}^2 = M_L^2(\chi_v^{-1} = 0)$, $g_0 = g_{SF}(\chi_v^{-1} = 0)$, and $g < 1/5$.

**Polarization of magnetoresistive manganites.**

Now we focus on the polarization of the ground state of manganites defined as $P = M/M_0$, where $M$ is the spontaneous magnetization given by the magnetic equation of state $\partial F / \partial M = 0$ given by

$$M^2 = -(1-5g)(\chi_0 \gamma)^{-1} - \frac{5}{3} M_{L0}^2 \tag{9}$$

and $M_0 = (-\chi_0 \gamma_0)^{-1/2}$ is the magnetization of totally polarized half-metallic ferromagnet given by the DE model. For the polarization we then have

$$P = (1 - \frac{5}{3} \frac{M_{L0}^2}{M_0^2})^{1/2}, \tag{10}$$

where we assumed that the renormalized anharmonicity parameter $g \ll 1$ is small and neglect the difference between the coupling parameters $\gamma$ and $\gamma_0$.

For manganites La$_{1-x}$A$_x$MnO$_3$ (where A is a divalent ion) $M_0$ can be estimated as $\mu_B(4-x)$ per Mn atom. Due to the lack of neutron scattering data for manganites we shall roughly estimate the value $M_{L0}^2$ as $3.2\mu_B^2$ taking it from the neutron scattering measurements in iron pnictides [11], which is also close to those found in other strongly correlated systems. As a result, for the $x = 0.3$ doping we found the ground state polarization $P \approx 78\%$. This rough estimate is surprisingly close to the experimentally measured value [3,4] $\sim 80\%$. Anyhow, this polarization is far from 100% predicted by the mean-field approximation of the DE model and allow for the low-frequency Stoner excitations which may essentially affect the magnon spectrum of manganites.

**Anomalous magnon damping and softening in manganites.**

In several manganite systems Pr$_{0.63}$Sr$_{0.37}$MnO$_3$, La$_{0.7}$Ca$_{0.3}$MnO$_3$, and Nd$_{0.7}$Sr$_{0.3}$MnO$_3$ magnons exhibit anomalies on approaching the Brillouin zone boundary [12]. At low temperatures magnon damping in these systems discontinuously increases near the wavevector $k_c \sim 0.3$ (in the reciprocal lattice units, r.l.u.). This anomalous jump in damping is accompanied by appreciate magnon softening near the Brillouin zone boundary, where magnon frequencies have different direction-dependent values. All mechanisms suggested so far, including four-magnon scattering, electron-phonon and electron-magnon interactions, and magnon scattering by orbital excitations [12] fail to account for this phenomenon [13]

The obvious reason for the abrupt increase of magnon damping may be related to the intersection of the magnon dispersion curve with the continuum of Stoner excitations leading to the strong Landau damping. Following the ideas of our paper [13] to describe magnon anomalies in manganites we present a simple phenomenological model consisting of two coupled fluids describing two coupled dynamical components of the system. The first component accounts for a ferromagnetic Fermi-liquid and the second one is related to a non-Fermi-liquid component of the magnetization.

Both components are characterized by the "partial" dynamical susceptibilities $\chi_{1,2}(\mathbf{k},\omega)$ and describing their linear response to the transverse magnetic field. Here we use a minimal description in order to explain qualitatively anomalies of the magnon spectrum of manganites and take the partial susceptibilities in the following simple form

$$\chi_1(\mathbf{k},\omega) = \chi_1 \frac{\omega_0(\mathbf{k})}{\omega_0(\mathbf{k}) - \omega - i\Gamma(\mathbf{k},\omega)},$$
$$\chi_2(\mathbf{k},\omega) = \frac{\chi_2}{[1+(\xi \mathbf{k})^2]}. \tag{11}$$

Here $\omega_0(\mathbf{k})$ is the frequency of "bare" optical magnons not vanishing in the long wavelength limit $(\omega_0(\mathbf{k}=0) = \omega_0 \neq 0)$, $\xi$ is the correlation length. The term $\Gamma(\mathbf{k},\omega) = \theta[\omega - \omega_S(\mathbf{k})]\omega\omega_0(\mathbf{k})/\omega_{fl}$ accounts for the Landau damping in the Stoner continuum, $\omega_{fl}$ is the characteristic frequency of transverse SF, the function $\theta[\omega - \omega_S(\mathbf{k})]$ is unity inside the Stoner continuum (when $\omega > \omega_S(\mathbf{k})$) and zero otherwise, and $\omega_S(\mathbf{k})$ is the lower boundary of the Stoner continuum.

The generalized susceptibility for the coupled fluids in the mean-field approximation takes the form

$$\chi(\mathbf{k},\omega) = \frac{\chi_1(\mathbf{k},\omega) + \chi_2(\mathbf{k}) + 2\lambda\chi_1(\mathbf{k},\omega)\chi_2(\mathbf{k})}{1 - \lambda^2 \chi_1(\mathbf{k},\omega)\chi_2(\mathbf{k})}. \tag{12}$$

where the constant $\lambda = const. > 0$ accounts for coupling which we assume to be ferromagnetic.

It should be emphasized that the susceptibility (12) is not related to any microscopic Hamiltonian being a phenomenological alternative to micrscopic descriptions, rather useful for the present analysis of the anomalies of the magnon spectra in manganites.

Near the magnon dispersion the generalized susceptibility

$$\chi(\mathbf{k},\omega) \sim \chi(\mathbf{k})\frac{\omega_m(\mathbf{k})}{\omega_m(\mathbf{k})-\omega-i\tau^{-1}(\mathbf{k})}, \qquad (13)$$

has a pole at the magnon frequency

$$\omega_m(\mathbf{k}) = \omega_0(\mathbf{k})[1-\lambda^2\chi_1\chi_2(\mathbf{k})], \qquad (14)$$

defining the true normal mode of the two-fluid system. Here $\chi(\mathbf{k})$ is the static transverse susceptibility, $\tau^{-1}(\mathbf{k})=\Gamma(\mathbf{k},\omega_m(\mathbf{k}))$ describes damping of magnons in the Stoner continuum. The factor $[1-\lambda^2\chi_1\chi_2]=0$ vanishes in the long wavelength limit [13] leading to the gapless Goldstone character of the magnon spectrum (14) which with account of Eqs. 11 takes the form

$$\omega_m(\mathbf{k}) = \omega_0(\mathbf{k})\frac{(\xi\mathbf{k})^2}{1+(\xi\mathbf{k})^2}, \qquad (15)$$

The magnon frequency (15) quadratically depends on the wavevector in the long wavelength limit $((\xi\mathbf{k})^2 \ll 1)$ and softens near the Brillouin zone boundary $((\xi\mathbf{k})^2 \gg 1)$,

$$\omega_m(\mathbf{k}) \approx \begin{cases} \omega_0(\xi\mathbf{k})^2 = D\mathbf{k}^2 \sim \mathbf{k}^2, & (\xi\mathbf{k})^2 \ll 1, \\ \omega_0(\mathbf{k}_B), & (\xi\mathbf{k})^2 \gg 1, \end{cases} \quad \omega_0(\mathbf{k}_B) \approx \begin{cases} \omega_1, & (\xi\mathbf{k})^2 \gg 1, [001] \\ \omega_2, & (\xi\mathbf{k})^2 \gg 1, [110] \end{cases}. \quad (16)$$

Here $D$ is the magnon stiffness, $\omega_0 = \omega_0(\mathbf{k}=0)$, and $\omega_0(\mathbf{k}_B)$ are zone-boundary frequencies in two directions [001] and [110] used in experiments [12].

Now we apply these results to explain anomalies of the magnon spectra in $Pr_{0.63}Sr_{0.37}MnO_3$, $La_{0.7}Ca_{0.3}MnO_3$, and $Nd_{0.7}Sr_{0.3}MnO_3$ which look surprisingly similar [12]. First, using the measured magnon stiffness [12] $\hbar D = \hbar\omega_0\xi_2^2 \approx 165$ meVA$^2$ we estimate the correlation length $\xi_2 \approx 2.74$ A and the wavevector $k_s = \xi_2^{-1} \approx 0.2$ r.l.u. which agrees well with the measured vector marking softening of the magnon spectra. It is also close to the value $k_c \approx 0.3$ where the jump in magnon damping takes place. To minimize the number of the parameters of the model we assume the energy $\hbar\omega_0$ to be equal to its measured zone-boundary value $\hbar\omega_0 = \hbar\omega_1 \approx 22$ meV in the [001] direction and set $\hbar\omega_1 \approx 45$ meV in the [110] direction.

**Conclusions**

To conclude, we show that zero-point SF impose fundamental limitations on the phenomenon of half-metallicity in ferromagnetic metals, which can give rise to low-frequency Stoner excitations forbidden, e.g., in the DE model usually used to describe magnetoresistive manganites. Using a phenomenological two-fluid approach we present evidence that anomalous magnon damping and softening in a series of manganites strongly suggest the presence there of low-frequency Stoner excitations breaking down their possible half-metallic character.

**Acknowledgements**

This work was supported by the State Atomic Energy Corporation of Russia "ROSATOM".

\* *E-mail address*: asolontsov@mail.ru.